\documentstyle[twoside,fleqn,espcrc2]{article}

\newcommand{\nc}{\newcommand}
\newcommand{\rnc}{\renewcommand}

\def\={\ =\ }
\def\+{\ +\ }
\def\-{\ -\ }

\nc{\Tr}{{\rm Tr\,}}
\nc{\rome}{{\rm Roma}}
\nc{\ie}{{\em i.e.}}
\nc{\eg}{{\em e.g.}}
\nc{\etal}{{\em et al.}}
\nc{\calF}{{\cal F}}
\nc{\calL}{{\cal L}}
\nc{\calS}{{\cal S}}
\nc{\calO}{{\cal O}}
\nc{\kapbar}{\bar{\kappa}}
\nc{\uidot}{\dot u}
\nc{\uiidot}{\ddot u}
\nc{\uiiidot}{\stackrel{\ldots}{u}}
\nc{\psii}{\psi^{(1)}}
\nc{\psibar}{\overline\psi}
\nc{\chibar}{\overline\chi}
\nc{\psiibar}{\overline\psi^{(1)}}
\nc{\psibari}{\overline\psi^{(1)}}

\rnc{\topfraction}{1.0}
\rnc{\bottomfraction}{1.0}
\rnc{\textfraction}{0.0}

\nc{\qq}{\P}
\nc{\qqc}{[\P]}

\nc{\rng}{\rangle}
\nc{\lng}{\langle}
\nc{\rcite}{ref.\ \cite}
\nc{\ba}{\begin{array}}
\nc{\ea}{\end{array}}
\nc{\lb}{\left(}
\nc{\rb}{\right)}
\nc{\qrt}{\frac{1}{4}}

\nc{\al}{\alpha}
\nc{\bt}{\beta}
\nc{\gm}{\gamma}
\nc{\dl}{\delta}
\nc{\ep}{\epsilon}
\nc{\varep}{\varepsilon}
\nc{\zt}{\zeta}
\nc{\et}{\eta}
\nc{\th}{\theta}
\nc{\kp}{\kappa}
\nc{\lm}{\lambda}
\nc{\rh}{\rho}
\nc{\sg}{\sigma}
\nc{\ta}{\tau}
\nc{\ph}{\phi}
\nc{\vr}{\varphi}
\nc{\ch}{\chi}
\nc{\ps}{\psi}
\nc{\om}{\omega}
\nc{\noi}{\noindent}
\nc{\half}{\frac{1}{2}}
\nc{\rr}[1]{$^{#1}$}
\nc{\rf}[1]{(\ref{#1})}
\nc{\rfs}[2]{(\ref{#1},\ref{#2})}
\nc{\smgr}{\stackrel{\textstyle <}{>}}
\nc{\grsm}{\stackrel{\textstyle >}{<}}
\nc{\aleq}{\mbox{}_{\textstyle \sim}^{\textstyle < }}
\nc{\ageq}{\mbox{}_{\textstyle \sim}^{\textstyle > }}
\nc{\ra}{\rightarrow}
\nc{\lra}{\leftrightarrow}
\nc{\be}{\begin{equation}}
\nc{\ee}{\end{equation}}
\nc{\bea}{\begin{eqnarray}}
\nc{\eea}{\end{eqnarray}}
\nc{\eqrf}{eq.\ \rf}
\nc{\erf}{{\rm erf}}

\nc{\ap}[1]{Ann.\ Phys.~#1\ }
\nc{\app}[1]{Acta Physica Polonica~#1\ }
\nc{\arnps}[1]{Ann.\ Rev.\ Nucl.\ Part.\ Sci.~#1\ }
\nc{\cmp}[1]{Commun.\ Math.\ Phys.~#1\ }
\nc{\cpc}[1]{Comput.\ Phys.\ Commun.~#1\ }
\nc{\jetp}[1]{Soviet Physics JETP~#1\ }
\nc{\jpa}[1]{J.\ Phys.\ A~#1\ } 
\nc{\jpg}[1]{J.\ Phys.\ G~#1\ } 
\nc{\mpla}[1]{Mod.\ Phys.\ Lett.~A#1\ }
\nc{\npa}[1]{Nucl.\ Phys.~A#1\ }
\nc{\npb}[1]{Nucl.\ Phys.~B#1\ }
\nc{\nproc}[1]{Nucl.\ Phys.~B (Proc.\ Suppl.)~#1\ }
\nc{\pla}[1]{Phys.\ Lett.~#1A\ }
\nc{\plb}[1]{Phys.\ Lett.~#1B\ }
\nc{\pr}[1]{Phys.\ Rep.~#1\ }
\nc{\pra}[1]{Phys.\ Rev.~A#1\ }
\nc{\prb}[1]{Phys.\ Rev.~B#1\ }
\nc{\prc}[1]{Phys.\ Rev.~C#1\ }
\nc{\prd}[1]{Phys.\ Rev.~D#1\ }
\nc{\pre}[1]{Phys.\ Rev.~E#1\ }
\nc{\prep}[1]{Phys.\ Rep.~#1\ }
\nc{\prev}[1]{Phys.\ Rev.~#1\ }
\nc{\prl}[1]{Phys.\ Rev.\ Lett.~#1\ }
\nc{\procroy}[1]{Proc.\ Roy.\ Soc.~#1\ }
\nc{\ptp}[1]{Prog.\ Theor.\ Phys.~#1\ }
\nc{\rmp}[1]{Rev.\ Mod.\ Phys.~#1\ }
\nc{\rpp}[1]{Rep.\ Prog.\ Phys.~#1\ }
\nc{\sjnp}[1]{Sov.\ J.\ Nucl.\ Phys.~#1\ }

\newcommand{\AmS}{{\protect\the\textfont2
  A\kern-.1667em\lower.5ex\hbox{M}\kern-.125emS}}

\title{Three topics in the Schwinger model\thanks{Contribution to Lattice '97
       by A.J. van der Sijs ({\tt arjan@scsc.ethz.ch}).  % \protect\\
       ETH-SCSC preprint {\bf TR-97-12}.}}

\author{Ph.\ de Forcrand,\address{Swiss Center for Scientific Computing,
        ETH-Z\"urich, ETH-Zentrum, CH-8092 Z\"urich, Switzerland}
        J.E. Hetrick,\address{Physics Department, University of the Pacific,
        Stockton, CA 95211, USA}
        T. Takaishi \address{Hiroshima University of Economics,
             Hiroshima,
             Japan 731-01}
        and
        A.J.\ van der Sijs$^{\rm \,a}$
       }

\begin{document}

\input epsf
\epsfverbosetrue

\begin{abstract}
1.\ We compare Monte Carlo results with analytic predictions for the fermion
condensate, in the massive one-flavour Schwinger model.
2.\ We illustrate on the Schwinger model how to facilitate transitions between
topological sectors by a simple reweighting method.
3.\ We discuss exact, non-perturbative improvement of the gauge sector.
\end{abstract}

\maketitle

\section{FERMION CONDENSATE}
In the massless (one-flavour) Schwinger model, the fermion condensate is
\be
\langle -\bar\psi \psi \rangle / \mu = e^\gamma/2\pi = 0.283466
 , \label{eq101}
\ee
in units of $\mu = g/\sqrt{\pi}$.
For non-zero fermion mass $m$, no exact expression is known, but there are
some results from analytical approximations and lattice simulations
\cite{cond}.
No clear agreement has been reached, though.

Here we report on a simulation using the $R$-algorithm \cite{R}
with staggered fermions,
in which extrapolations to zero step-size, to zero lattice spacing
and to infinite volume are performed.
We focus on two values of $m/\mu$: 0.33 and 1.00.
\begin{figure}[bt]
\centerline{
\epsfxsize=\columnwidth
\epsfbox{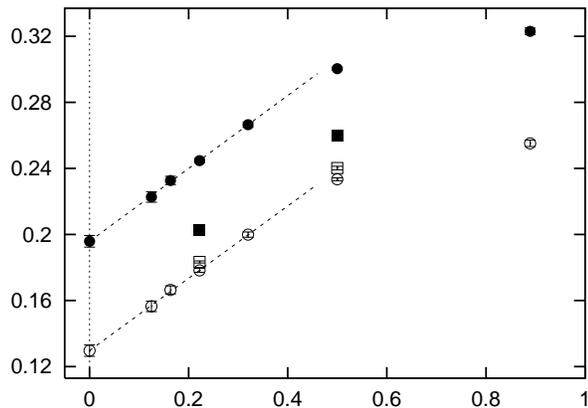}
}
\caption{Fermion condensate against $1/\beta$, with extrapolation, for fixed
physical volume and fermion mass.  The two sets of points correspond to
subtraction
of the free condensate in the same finite volume (solid circles) or in 
infinite volume (open circles).
The four squares correspond to 
$2\times 2$ times as large volumes.}
\vspace*{-1mm}
\label{fig101}
\end{figure}

Fig.\ \ref{fig101} shows the approach to the continuum limit for a given
physical volume: the condensate is plotted against $1/\beta$ for a series
of simulations on $(L/a)\times (2L/a)$ lattices ($L/a = 6,8,\ldots,16$),
keeping $am\sqrt{\beta\pi} = 0.33$ and $(L/a)/\sqrt{\beta} = 4\sqrt{2}$ fixed.
The condensate for the free theory is subtracted (in two different ways,
see caption)
in order to cancel the UV divergence present in the massive case.
A linear extrapolation to the continuum limit, accounting for
$\calO(1/\beta)\sim \calO(a^2)$ perturbative scaling violations,
works well and is seen to be very important, given the large effect.
After additional extrapolations to infinite volume and zero MD step-size,
we find for the (subtracted) condensate:
\be
\langle -\bar\psi \psi \rangle_{\rm sub} / \mu = 0.141(5)
    \ \ \ \ (m/\mu = 0.33)
 , \label{eq102}
\ee
and similarly for the larger mass:
\be
\langle -\bar\psi \psi \rangle_{\rm sub} / \mu = 0.084(3)
    \ \ \ \ (m/\mu = 1.00)
 . \label{eq103}
\ee

\begin{figure}[t]
\centerline{
\epsfxsize=\columnwidth
\epsfbox{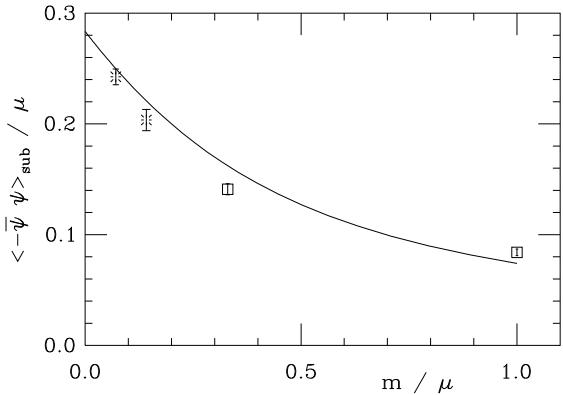}
}
\caption{The fermion condensate in infinite volume, in the continuum limit
(squares) compared with the analytical small-$m$ result of
Ref.\ \protect\cite{hosotani} (solid curve).  The two asterisks correspond to
our reanalysis of data from Ref.\ \protect\cite{azcoiti}.}
\label{fig102}
\end{figure}
In fig.\ \ref{fig102} these data are compared with a recent analytical result
by Hosotani\footnote{We thank Y. Hosotani for discussions on this topic.}
\cite{hosotani}, expected to hold in the small-$m$ region.
For comparison, we have also plotted our reanalysis ({\em i.e.,} subtraction
of the free condensate and subsequent linear extrapolation to $1/\beta=0$)
of data from Ref.\ \protect\cite{azcoiti} for the non-compact model on
a $64\,\times\,64$ lattice.

\section{SAMPLING TOPOLOGICAL\protect\\ SECTORS AT HIGH $\beta$}
A correct sampling of the different topological sectors
is essential for a correct determination of quantities like the
fermionic condensate in the Schwinger model.
In the usual importance sampling algorithms, however,
transitions between sectors are suppressed by a factor of order
$\exp(2\beta)$ in the continuum limit, implying {\em exponential\/}
critical slowing down for the topological autocorrelation time.
\begin{figure}[bt]
\centerline{
\epsfxsize=\columnwidth
\epsfbox{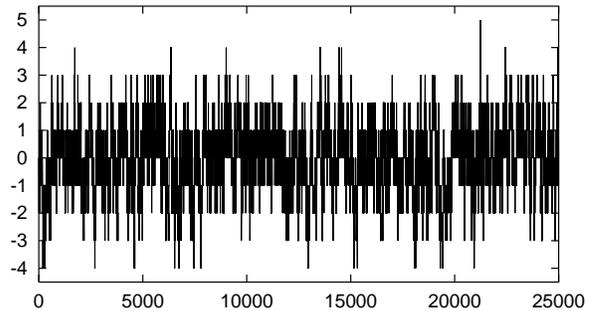}
}
\caption{
Topological charge history (in units of M.D. time) for a simulation on an
$18\,\times\,36$ lattice at $\beta=10.125$, using reweighting.
For comparison, without reweighting we observed only {\em one\/} transition
in 200\,000 units of M.D. time.}
\vspace*{-4mm}
\label{fig21}
\end{figure}

We propose the following reweighting technique to overcome this problem.
We add to the action the quantity
\be
\Delta S = - \sum_x \alpha \exp \left[ - \frac{(\theta_{\mu\nu}(x) - \pi)^2}
   {2\theta_0^2} \right] \, \delta_{x,x_0}
 , \label{eq201}
\ee
and compensate for this by including a factor
$\exp(\beta\Delta S)$ in the observables.
Here $x_0$ is the site at which the plaquette angle $\theta_{\mu\nu}$ is
closest to $\pi$ of all the plaquettes in the configuration under
consideration.
This procedure increases the topological transition rate by strongly
enhancing the probability that the plaquette at $x_0$ goes through $\pi$.
Note that $x_0$ may vary
from one configuration to the next, but the essential point is that
the prescription assigns a unique action to each configuration.
It can be shown that this leads to reversible molecular dynamics, ensuring
detailed balance.
The computational overhead is negligible.

The example in Fig.\ \ref{fig21} shows that this method drastically
improves sampling of the topological sectors (ergodicity).

The parameters $\alpha$ and $\theta_0$ can be used for optimization.
For example, they can be adjusted such that the distribution (histogram)
of $\theta_{\mu\nu}(x_0)$ is roughly flat over most of the interval
$[-\pi,\pi]$.
On large lattices, $\theta_{\mu\nu}(x_0)$ gets closer to $\pi$, such
that smaller $\alpha$ and $\theta_0$ suffice for the same transition rate,
and the bias due to Eq.\ \rf{eq201} gets smaller.

This reweighting procedure enhances a local lattice artefact, which acts
like a ``saddle point'' between
topological sectors in the lattice formulation; in the continuum theory,
these sectors are completely disconnected.
This observation may guide a simple generalization to non-abelian
gauge theories in 4 dimensions.

\section{NON-PERTURBATIVE IMPROVEMENT}
Two-dimensional U(1) lattice gauge theory with the standard plaquette action
is exactly solvable.
In particular, in the infinite-volume limit the expectation value of any
rectangular Wilson loop of charge $q$
satisfies an exact area law with string tension
\bea
a^2\sigma_q(\beta) &=& \ln \frac{I_0(\beta)}{I_q(\beta)} \label{eq301} \\
          &=& q^2 \left(\frac1{2\beta} + \frac1{4\beta^2}\right) +
      \calO\left(\frac1{\beta^3}\right)
 , \label{eq302}
\eea
where $I_{0,q}(\beta)$ are the Bessel functions.
For the non-compact action, on the other hand (cf.\ also the Manton action),
one finds $\sigma_q = q^2/2\beta$ (``perturbative scaling'') exactly.
We use this observation as a guideline for non-perturbative
improvement\footnote{This is different from the usual improvement where the
expansion of the plaquette angle in terms of the continuum gauge field plays
an essential role.}
of the compact plaquette action:
tune the coefficient of the improvement term such that the string tension
equals $q^2/2\beta$.

We consider the addition of an adjoint (squared plaquette) term to
the standard action:
\be
\Delta S = \sum_p c(\beta) \left[ \frac{1- \cos \theta_p}3 -
   \frac{1- \cos 2\theta_p}{12} \right] .
 \label{eq303}
\ee
In this case, the string tension is given by Eq.\ \rf{eq301} with $I_n$
replaced with $\lambda_n$, defined as
\be
\lambda_n(\beta_f,\beta_a) =
\int_{-\pi}^{\pi} \! \frac{d\theta}{2\pi} \;
\cos (n\theta) \, e^{\beta_f \cos \theta + \beta_a \cos 2\theta} ,
 \label{eq304}
\ee
with $\beta_f = \beta\,(1+c(\beta)/3)$, $\beta_a = -\beta\, c(\beta)/12$.
The $\lambda_n$ are calculated numerically.

\begin{figure}[tb]
\centerline{
\epsfxsize=\columnwidth
\epsfbox{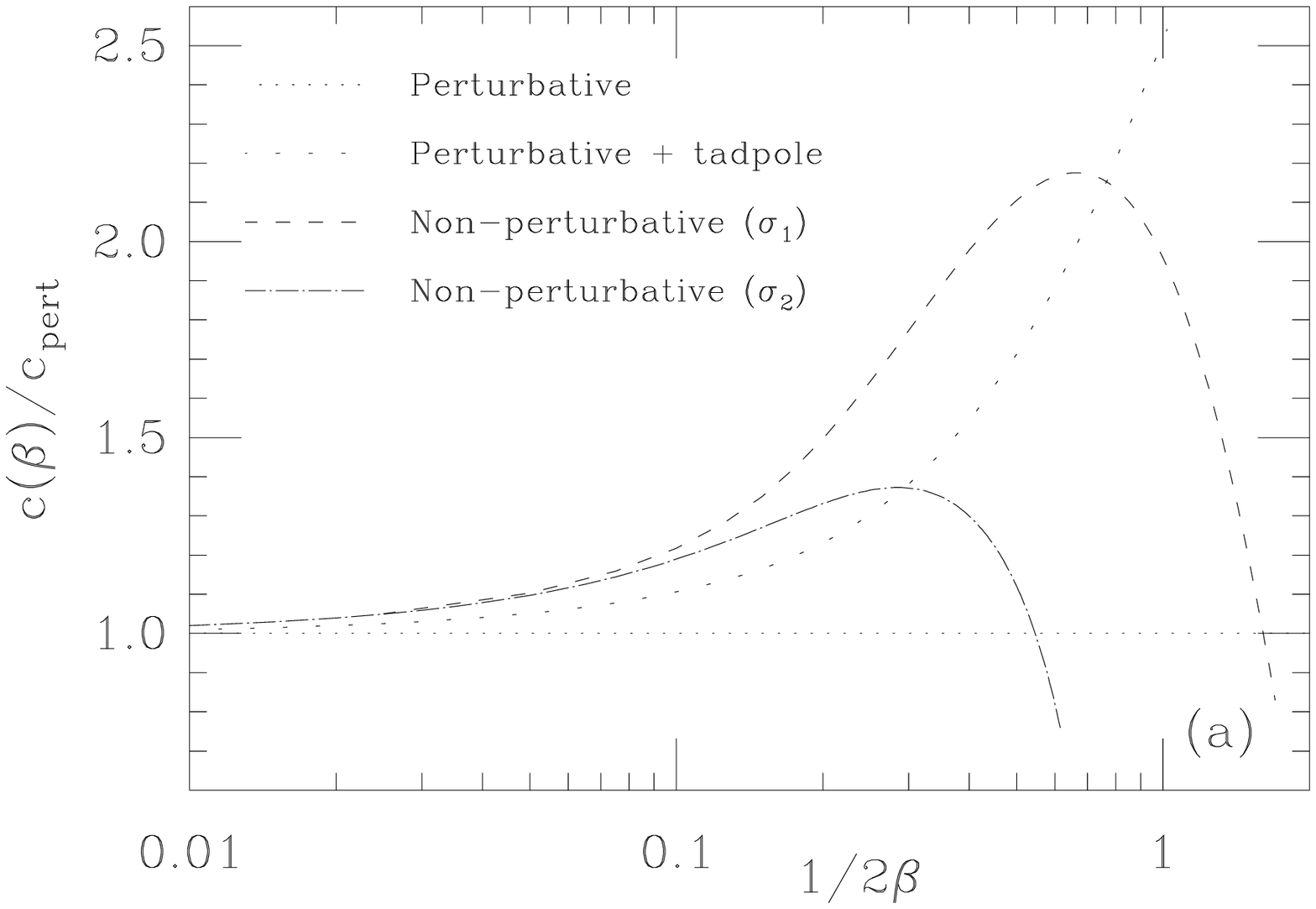}
}
\centerline{
\epsfxsize=\columnwidth
\epsfbox{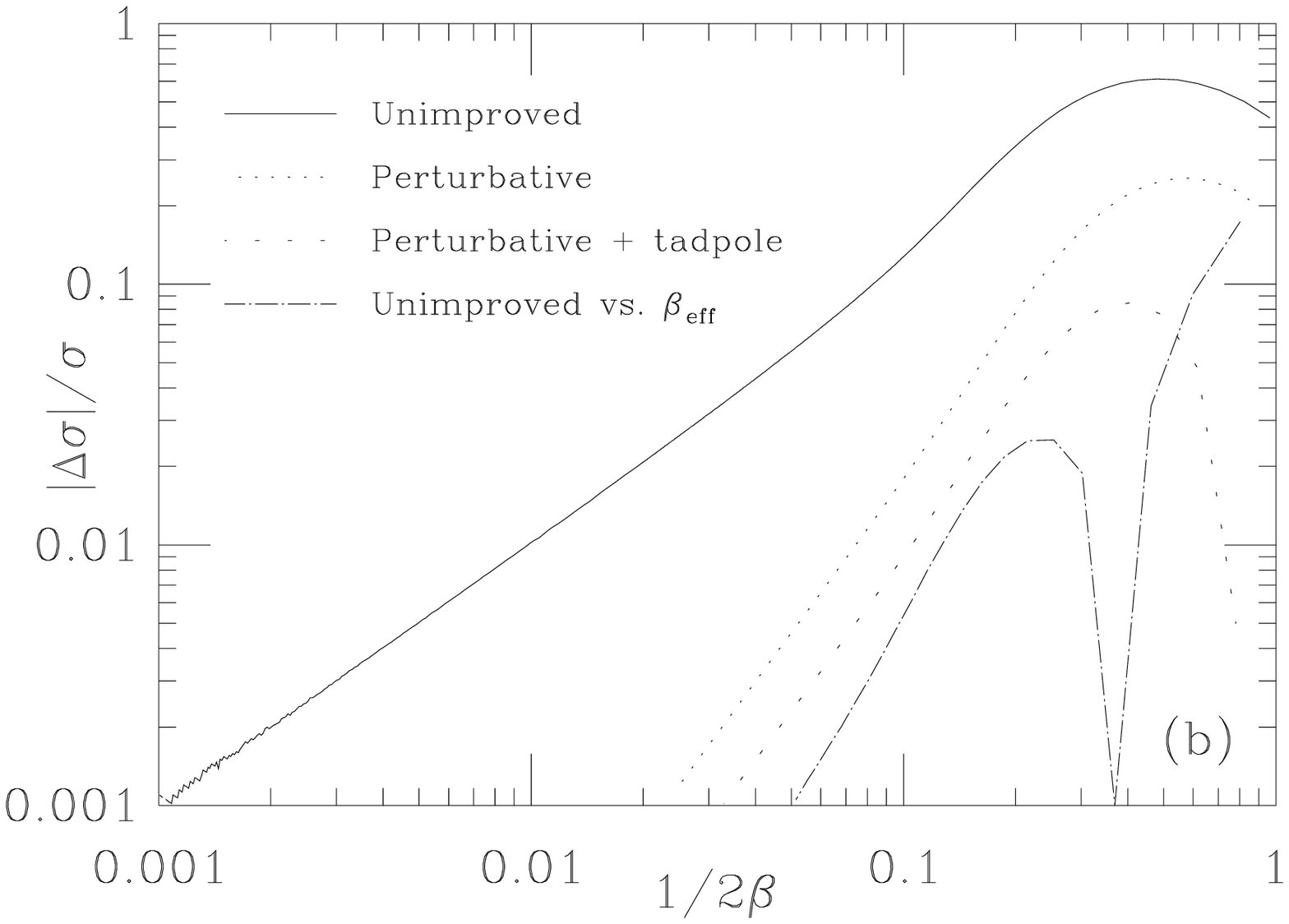}
}
\caption{
$c(\beta)$ for the improvement schemes described in the text ($a$)
and relative deviation of $a^2\sigma_1$ from the value $1/2\beta$ ($b$).
For the dash-dotted curve in $(b)$, the horizontal axis
is $1/2\beta_{\rm eff}$.
}
\label{fig31}
\end{figure}

In Fig.\ \ref{fig31} we compare various choices for $c(\beta)$.
By taking $c \equiv 1$, the $\theta_p^4$ term in the action, and hence the
$\beta^{-2}$ term in $a^2\sigma_q$ \rf{eq302}, are cancelled (perturbative
improvement): the slope of the relative ``error'' in the string tension
(fig.\ \ref{fig31}$b$) increases from 1 to 2.  The error
is seen to be further reduced when this coefficient is tadpole-improved,
by taking $c(\beta) = 1/\langle \Box (\beta)\rangle$ self-consistently.
Non-perturbative improvement amounts to tuning $c(\beta)$
such that $a^2\sigma_q = q^2/2\beta$ for all $\beta$ (for a given $q$).
Note that a very good result is obtained by simply using the unimproved
($c\equiv 0$) Wilson action but expressing the results in terms of the
effective coupling $\beta_{\rm eff} = \beta \,\langle \Box (\beta)\rangle$
suggested by the tadpole scheme.

\end{document}